# EMÍLIAS PODCAST - MULHERES NA COMPUTAÇÃO: AMPLIANDO HORIZONTES E INSPIRANDO CARREIRAS EM STEM

Nathálya Chaves Dos Santos[1], Adolfo Gustavo Serra Seca Neto[2]

**RESUMO:** No dia 3 de outubro de 2024, o "Emílias Podcast - Mulheres na Computação" completa 5 anos de existência, destacando-se como uma plataforma que promove a participação das mulheres na STEM (acrônimo em inglês para "ciência, tecnologia, engenharia e matemática"). O podcast visa abrir espaço para mulheres da área da computação e áreas correlatas compartilharem suas experiências e divulgarem as diversas oportunidades na Tecnologia da Informação e Comunicação (TIC). A metodologia utilizada incluiu uma pesquisa de feedback com entrevistadas, realizada via Google Forms, para avaliar a experiência das participantes e verificar se recomendariam o podcast. Além disso, analisamos dados de audiência, que mostraram um crescimento consistente ao longo dos cinco anos. Os resultados mostraram que 100% das entrevistadas recomendariam o "Emílias Podcast", refletindo a alta satisfação com o projeto. A avaliação média da experiência de participação foi de 4,7 em uma escala de 1 a 5, destacando aspectos positivos como a qualidade do roteiro, condução da entrevista e oportunidade de networking. Os dados de audiência também evidenciam o impacto do podcast: com mais de 10.000 downloads e reproduções acumuladas, é ouvido majoritariamente por pessoas entre 23 e 44 anos, com um público 50,9% feminino, demonstrando sua relevância e alcance. Em conclusão, o feedback das entrevistadas e os dados de audiência reforçam o impacto positivo do podcast e seu papel crucial na inclusão de mulheres na tecnologia. Os resultados evidenciam a importância de divulgar a área e suas oportunidades, contribuindo para um futuro mais inclusivo e inspirador. A análise dos dados demonstra a eficácia do podcast em engajar e expandir sua audiência, consolidando-se como um exemplo significativo de impacto social na TIC.

**Palavras-chave:** feedback, mulheres, tecnologia.

## 1 INTRODUÇÃO

O "Emílias Podcast - Mulheres na Computação" foi lançado em 2019 como parte do programa de extensão "Emílias - Armação em Bits", com o objetivo de atrair meninas e mulheres para a área de computação. Ao longo dos últimos cinco anos, o podcast tem se consolidado como uma importante plataforma para amplificar as vozes femininas na

---

[1] Bolsista da Fundação Araucária. Universidade Tecnológica Federal do Paraná, Curitiba, Paraná, Brasil. E-mail: nathalyachaves@alunos.utfpr.edu.br . ID Lattes: 0231194738307666.

[2] Docente no Programa de Pós-Graduação em Computação Aplicada (PPGCA). Universidade Tecnológica Federal do Paraná, Curitiba, PR, Brasil. E-mail: adolfo@utfpr.edu.br . ID Lattes: 00711197152724492.

Tecnologia da Informação e Comunicação (TIC) e em áreas correlatas. Além de abrir espaço para discussões sobre a representatividade feminina, o podcast também tem se mostrado um agente transformador para inspirar jovens mulheres a considerarem carreiras em STEM (ciência, tecnologia, engenharia e matemática), ampliando seus horizontes profissionais.

Conforme relatado por autoras como Mendes (2022), a presença de mulheres em STEM é ainda um desafio global. Iniciativas como podcasts desempenham um papel fundamental ao criar espaços onde essas mulheres podem compartilhar experiências e inspirar novas gerações. O "Emílias Podcast" se destaca por abordar temas diversos, desde a representatividade feminina até as inovações tecnológicas.

Neste artigo, analisamos os dados de audiência e o feedback das entrevistadas para avaliar o impacto do podcast e sua contribuição para a inclusão de mulheres na computação. O objetivo principal é entender como o podcast tem contribuído para aumentar a visibilidade dessas mulheres e promover a diversidade na área.

## 2  METODOLOGIA

Para realizar esta análise, utilizamos duas fontes de dados principais:

1. Pesquisa de Feedback: Envíamos um questionário para algumas das mulheres que foram entrevistadas pelo podcast e tiveram episódios lançados em 2024 e/ou tiveram contato com alguma vertente do projeto neste mesmo ano. Após o envio do formulário obtivemos retorno de 7 entrevistadas, portanto, os dados apresentados neste artigo se baseiam nos resultados obtidos destas respostas. A pesquisa, conduzida via Google Forms, incluiu perguntas sobre a experiência participando do podcast e se as entrevistadas recomendariam o podcast, além de um espaço para comentários e sugestões.

2. Dados de Audiência: Coletamos os dados de audiência do podcast através da plataforma de hospedagem Spotify for Podcasters - para este artigo não contaremos com dados do YouTube, onde também publicamos nossos episódios - que incluem informações sobre o número de downloads, idade e gênero dos ouvintes. Alguns destes dados (idade e gênero) são exclusivamente coletados para ouvintes que escutam pelo aplicativo Spotify. Analisamos estas métricas para avaliar o crescimento da audiência ao longo dos cinco anos.

## 3 RESULTADOS E DISCUSSÃO

### 3.1. Feedback das Entrevistadas

Os resultados da pesquisa de feedback mostraram que 100% das entrevistadas recomendariam o "Emílias Podcast".

Gráfico 1 – Resultado de pesquisa de feedback

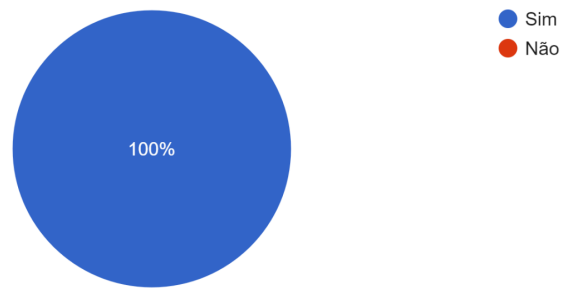

Fonte: Google Forms.

A avaliação média da experiência de participação foi de 4,7 em uma escala de 1 a 5.

Gráfico 2 – Resultado de pesquisa de feedback

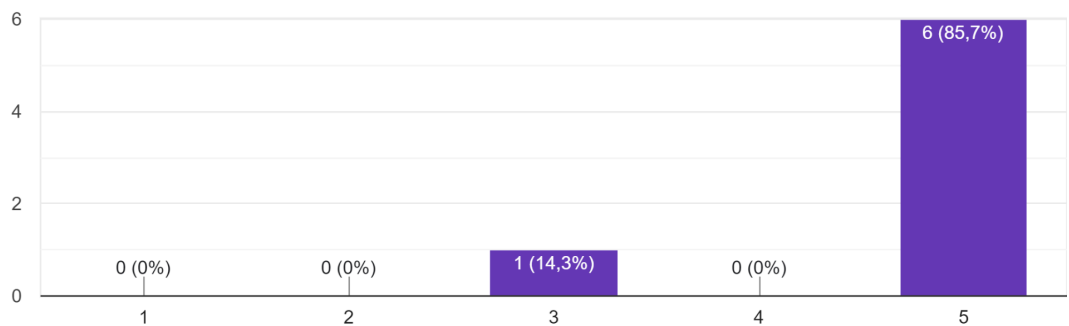

Fonte: Google Forms.

Como pontos fortes, as entrevistadas elogiaram a qualidade do roteiro, a fluidez das entrevistas e as oportunidades de networking oferecidas. Houve sugestões para a realização de episódios presenciais e eventos ao vivo, o que pode aumentar a interação com a audiência e o engajamento da comunidade acadêmica.

3.2. Dados de Audiência

Os dados de audiência revelaram um crescimento contínuo desde o lançamento do podcast. Até o momento, o "Emílias Podcast" acumula mais de 10.000 downloads e reproduções.

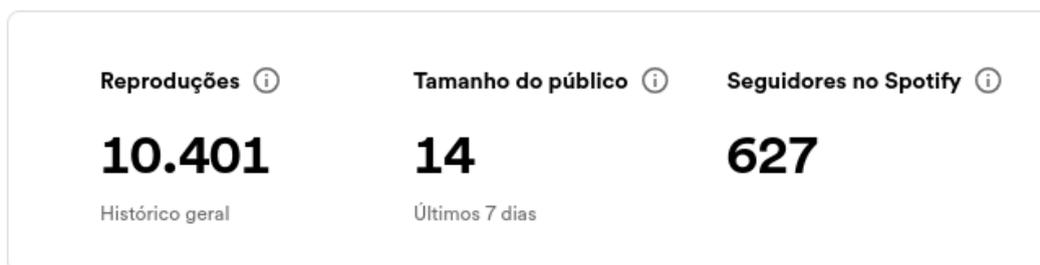

Figura 1: Dados de audiência. Em toda nossa história a soma de todos os nossos episódios foi baixada ou reproduzida lá 10.401 vezes. Deste total de 10.401, apenas 32,7% escuta no aplicativo Spotify, logado em sua conta em que há uma identificação de gênero. Ou seja, para os demais 67,3% não temos estes números. Fonte: Spotify

O público é composto majoritariamente por pessoas entre 23 e 44 anos, sendo 50,9% da audiência formada por mulheres.

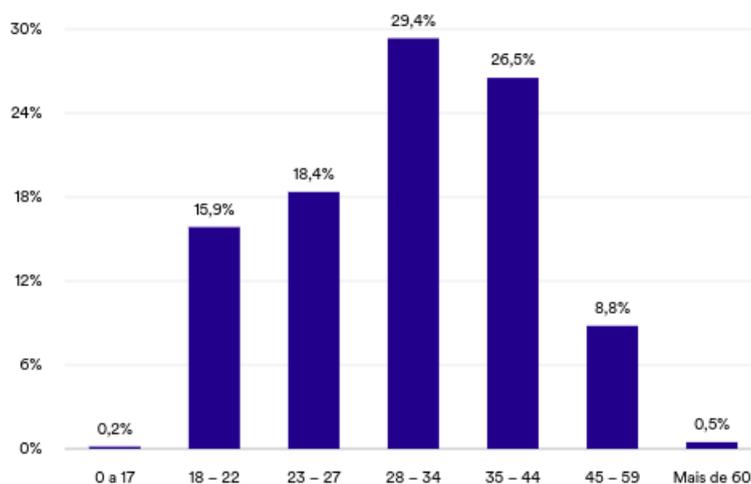

Figura 2: Dados de audiência. Faixa etária dos ouvintes em toda a história do podcast. Fonte: Spotify

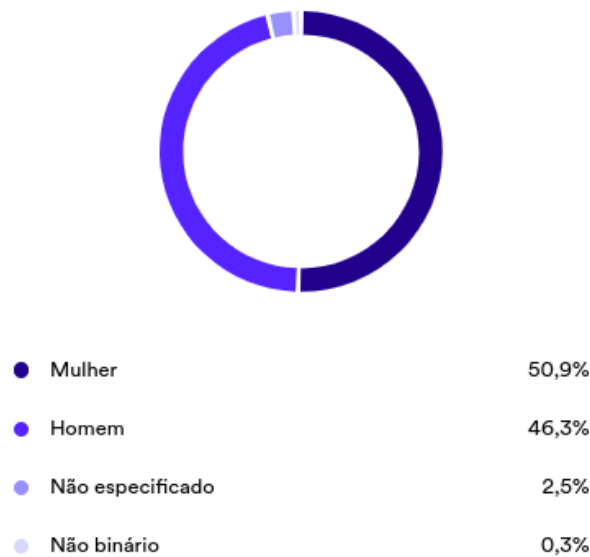

Figura 3: Dados de audiência. Gênero dos nossos ouvintes em toda história do podcast. Do total de 10.401, apenas 32,7% escuta no aplicativo Spotify, logado em sua conta em que há uma identificação de gênero. Ou seja, para os demais 67,3% não temos estes números. Fonte: Spotify

O crescimento contínuo da audiência mostra que o conteúdo tem ressoado positivamente entre os ouvintes. A predominância de mulheres na audiência reforça o impacto direto na comunidade feminina, mostrando que o podcast tem alcançado com sucesso seu público-alvo, promovendo representatividade e estimulando o interesse pela área de tecnologia, o que também pode ser evidenciado mediante feedbacks públicos de mulheres que adentraram na área da computação também por influência do "Emílias Podcast".

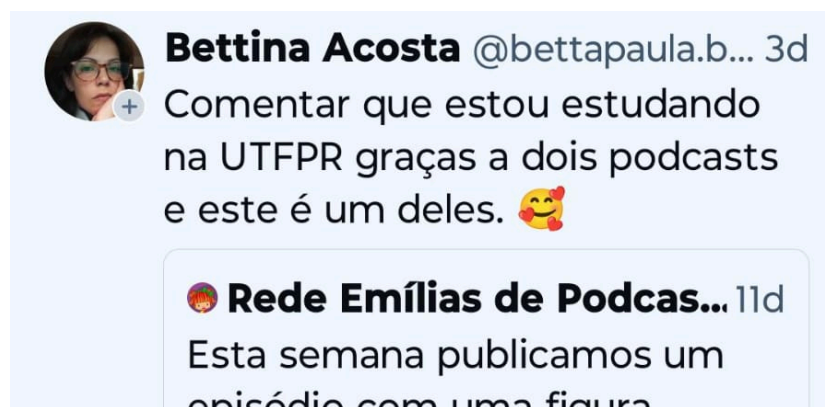

Figura 4: Feedback público. Impacto no objetivo de atrair mulheres para área da computação. Fonte: BlueSky

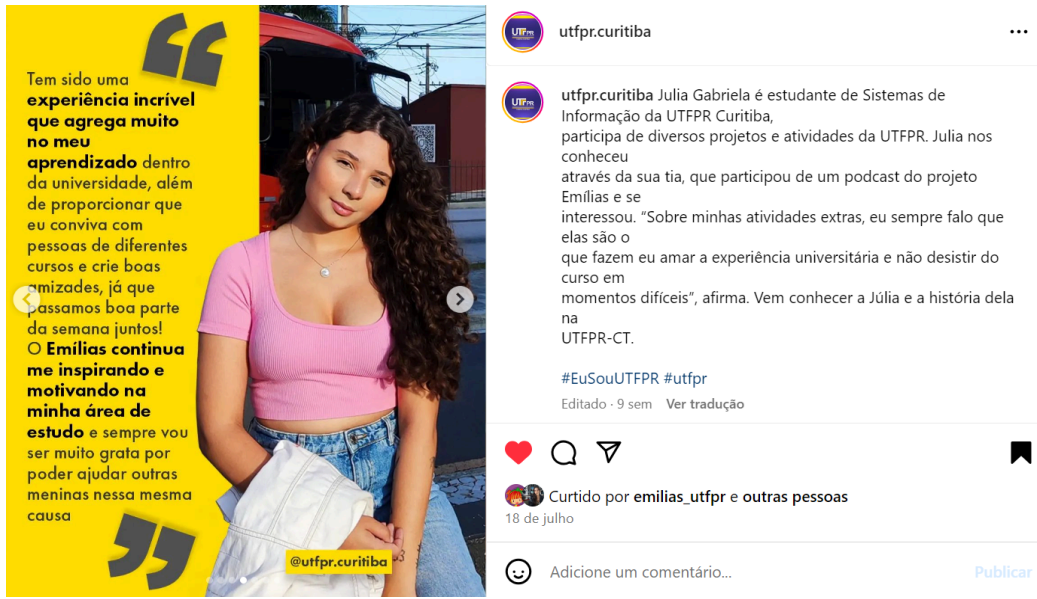

Figura 4: Feedback público. Impacto no objetivo de atrair mulheres para área da computação e papel do Emílias enquanto apoio durante o processo de formação como projeto de extensão. Fonte: Instagram UTFPR Curitiba (@utfpr.curitiba)

## 4 CONSIDERAÇÕES FINAIS

Após cinco anos de existência, o "Emílias Podcast - Mulheres na Computação" se estabeleceu como uma plataforma de impacto positivo na inclusão de mulheres na área de TIC.

Os resultados deste estudo evidenciam a importância de iniciativas como o "Emílias Podcast" na promoção de um ambiente mais inclusivo na tecnologia, inspirando novas gerações de mulheres a ingressarem e prosperarem na área de computação. Com sua capacidade de engajar e inspirar, o podcast tem influenciado diretamente a formação de novas gerações de mulheres que desejam construir suas carreiras em áreas relacionadas a STEM.

## AGRADECIMENTOS

## REFERÊNCIAS